\documentclass[preprint,flushrt]{aastex}

\begin{document}

\title{Unidentified Features in the Ultraviolet Spectrum of X Per\footnote{Based 
on observations made with the NASA/ESA {\it Hubble Space Telescope}, obtained from the 
Data Archive at the Space Telescope Science Institute, which is operated by the 
Association of Universities for Research in Astronomy, Inc., under NASA contract 
NAS 5-26555.}}

\author{Joshua D. Destree and Theodore P. Snow}

\affil{Center for Astrophysics and Space Astronomy}
\affil{Department of Astrophysical and Planetary Sciences, University of Colorado at Boulder}
\affil{Campus Box 389}
\affil{Boulder, CO 80309-0389}
\email{destree@casa.colorado.edu, tsnow@casa.colorado.edu}

\begin{abstract}
High-resolution ultraviolet spectra from the Space Telescope Imaging Spectrograph (STIS) 
were used to search for unidentified interstellar absorption features in the 
well studied sightline towards X Per (HD 24534).  The significance of features 
detected was determined from Gaussian fits to the data, as well as the features' 
persistence in multiple observations.  Fixed pattern noise characteristics were studied in 
STIS echelle data to distinguish between interstellar and instrumental features.
We report the detection of two unidentified features that stand out from 
the more common fixed pattern noise features.  Both features have depths of $>$ 3\%
of the continuum level making them very likely of interstellar origin. 
Lastly, we comment on possible carriers, and discuss future prospects for 
studying these and perhaps other unidentified lines in larger samples of sightlines.
\end{abstract}


\section{Introduction}
Over the last century, our knowledge of the 
components and structure of the diffuse interstellar medium 
(ISM) has progressed enormously.  We have now spectroscopically 
detected hundreds of different interstellar absorption and 
emission features throughout the electromagnetic spectrum,
leading to a deeper understanding of the physical and chemical composition of the ISM.

The majority of interstellar molecules have been detected only through 
infrared and millimeter emission spectra (Snow \& Bierbaum 2008).  
UV absorption line studies are of great value, though, as many 
atoms and molecules have ground state electronic transitions 
in the UV region of the electromagnetic spectrum.
Absorption lines give precise measures of column densities 
given sufficient spectral resolution and if the oscillator strength 
(f-value) of the transition is known -- 
unlike emission studies, absorption does not require a detailed 
excitation model to calculate column densities.
Measured column densities or equivalent widths can then be directly 
compared to other species observed in the spectra.

With several space-based observatories and sub-orbital experiments able to collect UV spectra 
over the past 40 years, much progress has been made in understanding the absorption spectra 
of many interstellar species.  Several molecules have been studied in the UV including 
H$_2$, HD, CO, C$_2$, CH, HCl, OH, CH$_2$, and CS (Snow \& Bierbaum 2008).
The many studies of such molecules, as well as atoms and ions in the interstellar medium,
have led to a greatly increased understanding of the chemistry and processes governing 
different regions of the ISM.  Models of interstellar clouds have become increasing
detailed with each passing decade.
Yet, many molecules with UV transitions have not been detected and
much could be gained from their study.  Polycyclic aromatic hydrocarbons (PAHs) for 
example have been detected in the IR through emission.  However, IR emission lines cannot 
tell us exactly which molecules exist in interstellar clouds; rather, they can only 
give us information on which vibrational modes the molecules have (Leger \& Puget 1984, and Draine 2003).

The energy released in IR emission is probably supplied by absorption of visible or UV 
photons (Draine 2003). Whereas the IR spectrum of various PAHs is not unique, their visible and UV spectra 
should be.  Thus, their detection in the UV might allow us to pinpoint which exact chemical structures 
are formed in abundance in interstellar clouds.

The rest of this paper, describing our search for unidentified interstellar
absorption features in the UV spectrum, is organized as follows.  In \S\ 2 we quickly review the methodology and results of
previous searches for unidentified lines in the UV and justify the current search which seeks
to use a different methodology to searching for unidentified features.  In \S\ 3 we describe 
the data and reduction, analysis techniques used, and challenges faced analyzing STIS echelle data.  
In \S\ 4 we describe the results of the current study, and in \S\ 5 we summarize our conclusions 
and discuss future research prospects.

\section{Previous studies and current methodology}
This work is by no means the first search for unidentified interstellar features in the UV.
Several studies have been performed searching for broad features in the extinction curves 
of reddened stars, in order to find a UV corollary to the many known visible diffuse interstellar bands (DIBs).
Many have proposed that large organic molecules such as PAHs could be viable DIB carriers 
(see Herbig, 1995, for a detailed review of proposed molecular carriers and Salama et al. 1996, 
and Salama 2007 for more recent developments).  Snow, York, \& Resnick (1977)
searched for diffuse bands in the UV using {\it Copernicus} data from lightly 
reddened sightlines (E(B-V) from 0.2 to 0.4).  They searched for features by making extinction curves 
over the region from 1114 \AA\ to 1450 \AA .  Initially a possible feature was suggested at 1416 \AA ;
however, this turned out to be most likely from a mismatch of stellar types when creating the extinction curve.
Next, Seab \& Snow (1985) searched through {\it IUE} data for diffuse bands between 1150 \AA\ and 2300 \AA .  
The study led to upper limits on the strengths of any features as 
none were detected across the UV band studied.  This study had 
an advantage over the previous {\it Copernicus} study in that more highly reddened sightlines were 
used (up to an E(B-V) of 1.13). Almost a decade later, a possible UV DIB was identified at 1369 \AA\
by Tripp, Cardelli, \& Savage (1994) using Goddard High Resolution Spectrograph (GHRS) data from 
the well studied sightline $\zeta$ Oph. This feature was later identified as most likely from CH 
(Watson 2001). A recent study by Clayton et al. (2003) 
looked at the extinction curves of some highly reddened sightlines (E(B-V) from 1.22 to 1.73) using low
resolution data from the Space Telescope Imaging Spectrograph (STIS).  These extinction curves were then compared 
to absorption spectra from a few small PAHs.  This study also found negative results and set a limit on broad
 (20 \AA\ wide) features.  All together, these studies have shown that strong, very broad features
like the 4428 \AA\ visible DIB are absent from the UV extinction curves.

The current study is substantially different from those done previously in its methodology and focus.
The previous studies almost exclusively looked for very broad features in low resolution data 
by making extinction curves in the UV.  Such studies overlook any features that are not extraordinarily
broad and are limited in detecting very weak features due to mismatched stellar lines.  
Many of the DIBs in the visible region (which may be features of neutral
PAHs, Salama 2007) are narrow, having widths $\leq$ 1 \AA .  UV features of similar energy width 
would be expected to be only several hundredths to a couple tenths of an Angstrom wide 
(Tripp, Cardelli, \& Savage 1994).  Recent lab work studying the spectra of some PAHs in the gas phase 
confirmed that features from neutral PAHs have narrower profiles (with full width half maxes [FWHMs] on the order of a few cm$^{-1}$, Salama 2007).
Small neutral PAHs are also expected to absorb in the UV (Salama et al. 1996).
Thus, there is good justification to look for sharper unidentified features than those in previous searches.  

The sightline to X Per (HD 24534) is, in many ways, ideal to search for unidentified features.  The signal-to-noise ratio (S/N) of
high resolution STIS data in this sightline is far better than most sightlines probing interstellar clouds.
X Per is also an ideal choice due to its moderate reddening (E(B-V) = 0.59), simple velocity structure, 
high molecular content, and fast stellar rotation rate ($V sin\ i$ = 191 km s$^{-1}$, Penny 1996).
Also, X Per is known to have many DIBs in the visible region including the family of DIBs related to C$_2$
(Thorburn et al. 2003).

\section{Data Analysis}
All data were downloaded preprocessed from the Multimission Archive at Space Telescope Science Institute.
All X Per datasets used were taken with high resolution echelle modes using the E140H grating.
Data from the STIS E140H grating have a spectral resolution of approximately 
114000 and a pixel spacing of about 5 to 7 m\AA .
Where multiple STIS exposures and observations existed, the calibrated wavelengths 
were used to align spectra, then exposures were coadded.  Using calibrated wavelengths, rather than 
correlating on sharp features, does not introduce significant errors in 
absorption profile shape as calibrated wavelengths for the STIS MAMA detectors 
are accurate to less than a pixel within an exposure and 1 pixel between exposures 
(STIS Instrument Handbook\footnote{\tt http://www.stsci.edu/hst/stis/documents/}).
Spectra of different spectral 
resolutions were not coadded.  Ten pixels at all order edges were excluded from the coadd as they were 
frequently discrepant.  Pixels with high data quality flags 
were also excluded from the coadd.  Echelle orders were individually scaled 
to match the local mean flux level of all observations in order to prevent 
false features from being created at order edges and in regions where pixels
were cut due to data quality flags.  Scaling to the mean flux level of all observations 
may affect the accuracy of the absolute flux level; however, scaling does not adversely 
affect the present study as the profile of features smaller than the order size 
should remain unchanged.

Once the final spectrum was created, the first step performed in analyzing the data
was a blind search for absorption features.  The data were analyzed by stepping through 
the spectrum three pixels at a time and fitting a single Gaussian profile 
with a low order polynomial background, then flagging any features with greater than 2$\sigma$ significance.  
Using a Gaussian profile does have a disadvantage in that it
assumes the profile is approximately symmetric, which may not be true for all absorption bands.
To perform any type of blind search, though, some profile must be assumed as a starting point, 
and Gaussian fits adequately assess the general strength and width of an absorption profile, 
even for relatively asymmetric bands.  A Gaussian profile was chosen for it is often a good 
approximation of the instrumental line spread function and innate absorption profile for weak lines, 
as well as for the relative ease of implementing a Gaussian profile.
This type of blind search also has the disadvantage that blended features will be difficult to 
distinguish from one another and may be flagged as one feature.  This was not a major concern, though,
as we were searching for clearly defined, unidentified features.
The data were fit in 40, 80, and 160 pixel pieces (0.24 \AA , 0.49 \AA , and 0.98 \AA\ sections respectively)
to detect relatively narrow, medium, and broad features. We used the non-linear least-squares curve 
fitting algorithm MPFIT, by C. Markwardt,\footnote{\tt http://cow.physics.wisc.edu/$\sim$craigm/idl/idl.html} 
to find the best fit parameter values.  MPFIT is a set of routines 
that uses the Levenberg-Marquardt technique to minimize the square 
of deviations between data and a user-defined model.  These routines 
are based upon the MINPACK-1 Fortran package by Mor\`{e} and 
collaborators.\footnote{\tt http://www.netlib.org/minpack}

Once the blind analysis was performed, we inspected each possible detection and 
eliminated those with poorly defined continua.  Well-defined continua were required to 
be reasonably fit by a low order polynomial, and the existence of a feature had to be due 
to what appeared to be absorption, rather than a sharp rise in the spectrum.
Unfortunately, this was the most subjective part of the selection process.
The remaining list of possible features was the starting point for our search for
unidentified features.

An absorption line being fit with a Gaussian 
profile to a certain significance is in no way strong enough criteria to claim an 
unidentified feature.  Unlike most other studies of the ISM where a set of specific known lines are measured,
the current study uses as much of the spectrum as possible to search for features.
Because we are not looking for a feature at a specific wavelength, the probability of finding any false feature 
that meets our selection criteria is much higher.  This fact is easy to see through a simple application of Bayes
theorem.  The probability that a feature is real given that it passes our criteria is given by:
\begin{equation}
P(Real|Pass) =  \frac{P(Pass|Real)\times P(Real)}{P(Pass|False)\times P(False) + P(Pass|Real)\times P(Real)}
\label{eqn_bayes}
\end{equation}
where $P(Real)$ is the probability that any one unidentifiable feature is a real interstellar feature, 
$P(False)$ is the probability that any one unidentifiable feature is not a real interstellar feature ($P(False) = 1-P(Real)$), 
and notation like 
$P(Pass|Real)$ is the probability that a feature will pass our criteria given that it is indeed a real feature.
To maximize $P(Real|Pass)$ we can either define parameter cutoffs to increase $P(Real)$ and decrease $P(False)$
or we can set criteria that have a high $P(Pass|Real)$ and a low $P(Pass|False)$.  

If false features (features originating from the random noise, stellar background, or fixed pattern noise)
significantly outnumber real unidentified features, as we expect in the current study, then our criteria
must be sufficiently strict so that the $P(Pass|False)$ is very small.  Many of the probabilities in this 
equation we do not know outright; however, if we estimate that false features outnumber real features 100 
to 1, then our selection criteria would need to reject 99.95\% of false features while passing 95\% of 
real features if we were to assign a 95\% confidence level on detecting any given unidentified interstellar feature 
(the ratio of 100 false features per single real unidentified feature is reasonable given the 
results of the current study).

The following sections outline our strategy to reject the various categories of false features.

\subsection{Rejecting Photon Noise Features}
Because of photon statistics, random features due to Poisson noise are inevitable, independent of detector quality.
However, these features are relatively easy to combat.  Because of its nature, photon noise will create very few 
features that will be fit to high significance with a single Gaussian.  We tested the occurrence of these random features by 
simulating data sets of a flat continuum with Poisson noise and scanning them for features.  
Features greater than 2$\sigma$ significant occurred 
at a rate of one per 2000 pixels, while features greater than 3$\sigma$ significant occurred at a rate of one per
$1\times10^6$ pixels.  Thus, we selected a minimum significance of 3$\sigma$, which makes the expected number of 
random features due to photon noise in the X Per dataset much less than one (making P(False) small for
photon noise features).

\subsection{Rejecting Stellar Features}
Stellar features are complicated as they are usually broad features that will be consistent from one exposure to the next.
Though the rotational velocity of X Per is approximately 190 km/s, the overlap of different stellar features can
create features in the spectrum that have widths somewhat less than the star's $V sin\ i$.  
To accurately identify a feature as interstellar, we must either 1) select features narrower than could be created by 
overlapping stellar lines, 2) remove the effects of stellar lines through modeling or a unreddened stellar standard, 
or 3) confirm the existence of the feature in multiple different sightlines to stars of different stellar types.

Due to the methodology of the current study, we chose to employ the first strategy, making a cutoff on the width of 
acceptable features.  Though removing stellar lines or performing a broader study of multiple targets are possible, such techniques
introduce complexities that were purposefully avoided in the current study.  Creating extinction curves through 
standard stars or modeling introduces false structure due to stellar mismatch or model inaccuracies.  The current 
study focuses on narrower and weaker features than previously sought in studies using extinction curves.  We 
did attempt to confirm the existence of any unidentified interstellar features, though this too has difficulties
as the data quality of the X Per dataset is significantly higher than most interstellar sightlines.  Nevertheless,
confirmation did prove to be helpful in some cases.  Certainly a more thorough search over many sightlines could
be performed and stacking different datasets might be very helpful in revealing unidentified features; such 
techniques are beyond the scope of the current study but will be discussed later in the paper.  One last 
technique worth mentioning to distinguish between stellar and interstellar lines is observing sightlines to 
spectroscopic binaries and making multiple observations of the star in different phases of its orbit.  
This is of course not a viable technique for X Per.

To reject stellar features in the current study, we then needed to set a cutoff on the maximum width of an interstellar feature.  
To determine how stringent a cutoff was necessary, we again modeled several synthetic datasets, this time using stellar 
models from the TLUSTY grid of O and B stars as background (Lanz \& Hubeny 2003 and 2007).  These stellar models were convolved with a 
standard stellar rotation profile with a $V sin\ i$ of 190 km/s.  The standard IDL routine 
LSF\_ROTATE\footnote{\tt http://idlastro.gsfc.nasa.gov/ftp/pro/astro/lsf\_rotate.pro} was used to create 
the stellar rotation profile.  Poisson noise was added to the stellar continuum and the synthetic datasets 
were scanned for absorption features. 
We find that a FWHM 
cutoff of $\le$ 30 km/s rejects 94\% of all detected stellar features while a cutoff of $\le$ 20 km/s rejects 99\% of
all detected stellar features.  Thus we choose a cutoff of 20 km/s 
(approximately 0.09 \AA\ across our dataset and about 10\% of the star's  $V sin\ i$)
to avoid claiming stellar features as interstellar.  This effectively makes the P(False) for accepting 
stellar features very low. With about 260 features flagged in 3000 \AA\ of synthetic data, the expected number 
of features passed in the spectrum of X Per is much less than one.

\subsection{Rejecting Fixed Pattern Noise Features}
\begin{figure}
\begin{center}
\scalebox{.5}[.5]{\includegraphics[angle=90]{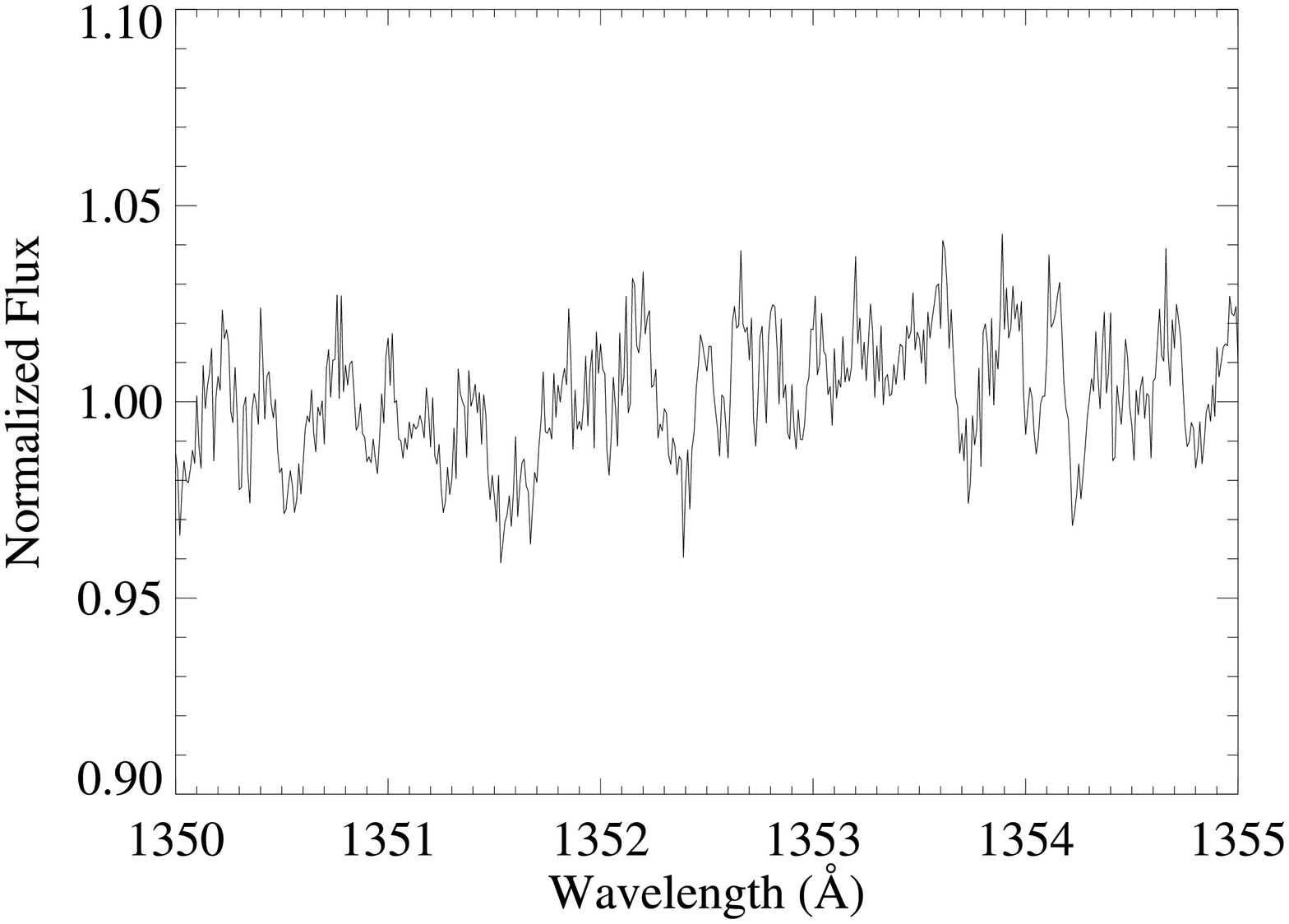}}
\caption{Plot of the difference spectrum between two overlapping X Per datasets.  All of the features in this plot are instrumental in origin -- fixed pattern noise features typically had depths of a few percent of the continuum. \label{fig_diffspec}}
\end{center}
\end{figure}

The most difficult types of features to reject in STIS echelle data were fixed pattern noise features.
The problems of fixed pattern noise were noticed in this sightline through comparisons of the edges 
of echelle orders, as well as comparing regions of overlap between observations with different central 
wavelengths.  Data from observations centered at 1343 \AA\ overlapped with data centered at 1416 \AA\ 
over a 120 \AA\ range (1320 to 1440).  We rebinned both of these data sets 
so that they had the same pixel spacing ($\Delta\lambda$ = 0.01 \AA ) and then divided the spectra from each other.
If the detector and calibration were perfect and the line of sight was not time dependent, we would expect the 
difference spectrum to be a flat line at 1 with noise due to photon statistics.
However, in comparing overlapping datasets we found that shallow ($\leq$ 3\% depth) features are rampant in 
the STIS echelle data (See Figure \ref{fig_diffspec}).  

\begin{figure}
\begin{center}
\scalebox{.5}[.5]{\includegraphics[angle=90]{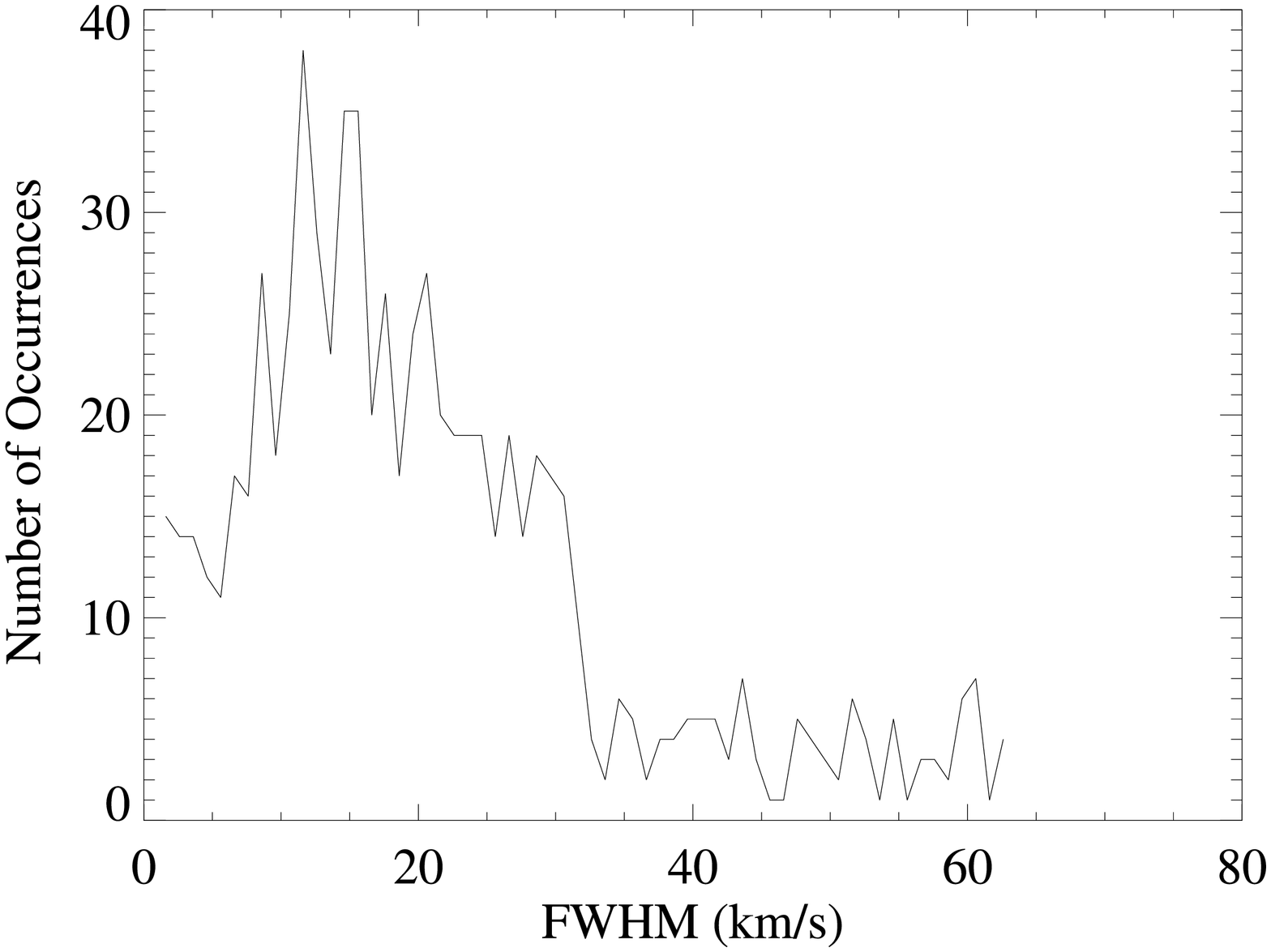}}
\scalebox{.5}[.5]{\includegraphics[angle=90]{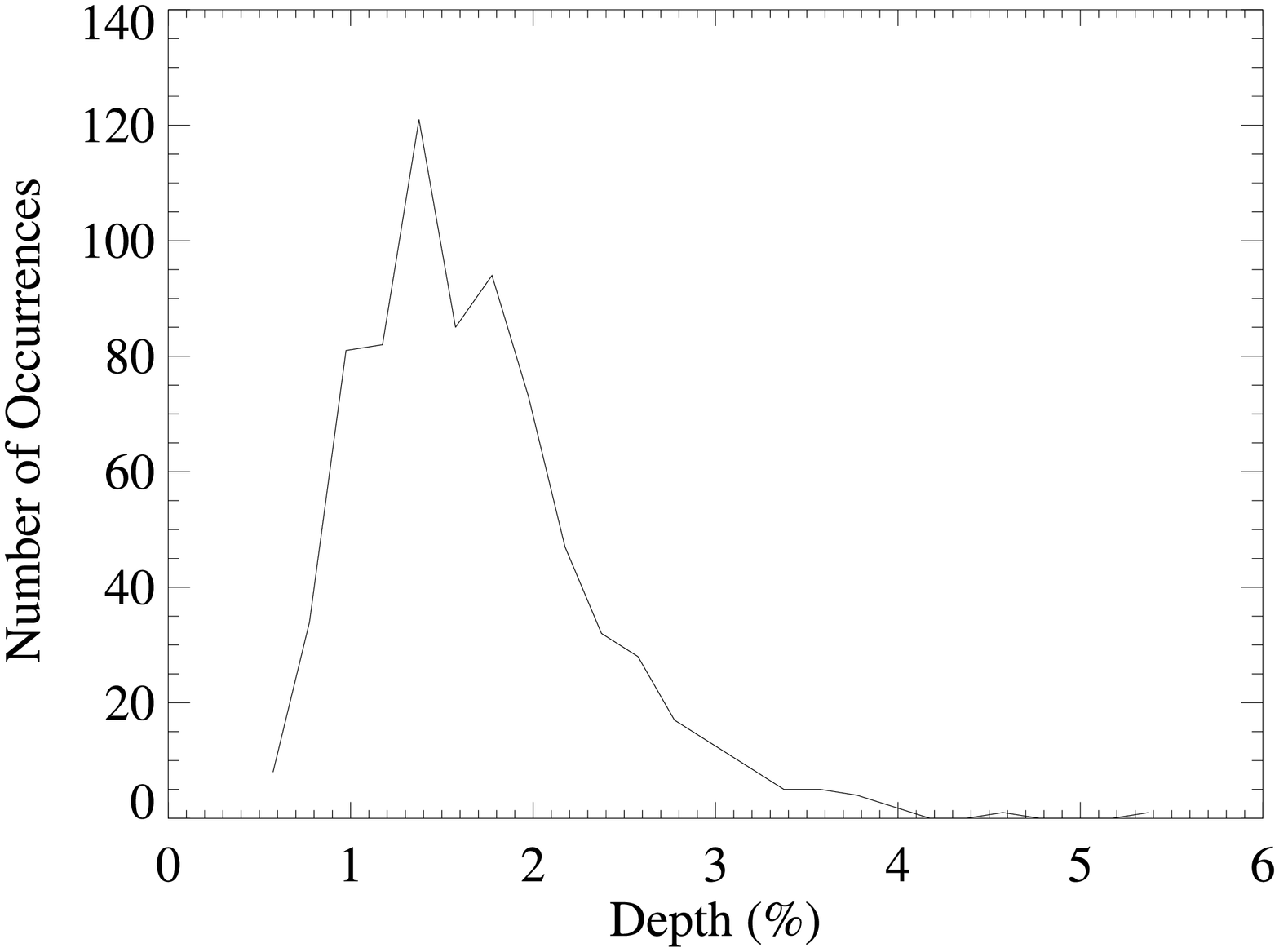}}

\caption{Histogram plots showing characteristics of fixed pattern noise in STIS echelle data.  The top plot shows typical FWHMs of instrumental features while the bottom plot shows typical depths (in percent of continuum level). \label{fig_fpchar}}
\end{center}
\end{figure}

To properly address whether these instrumental features could be distinguished from 
interstellar absorption, we need to know the characteristics of the features.
As little analysis has been done on describing instrumental noise at this level, 
we undertook a rough study of their characteristics. We first masked 
out all identifiable atomic or molecular lines (see Table \ref{tab_lines} for line 
lists used in identifying interstellar features).  We then scanned
the two different overlapping datasets in the region of 1320 \AA\ to 1440 \AA\ for features 
that were at least 2$\sigma$ significant.  In the 240 \AA\ of spectra scanned, we flagged 
742 features.  We can assume with good justification from the difference spectrum that the
vast majority of these features are instrumental in origin. Making histograms of the
detected features' characteristics gives us some basic information about them (see Figure
\ref{fig_fpchar}).  From these plots we can see that the vast majority
of detectable instrumental features have depths between 0.5\% and 3\% of the continuum level.
We also find that the majority of instrumental features detected by scanning the spectrum have
FWHMs of less than 35 km/s.  Correlating the detected instrumental features in wavelength, FWHM, and depth, 
we find that the feature characteristics have no measurable correlation to one another.
Thus, instrumental features were spread approximately uniformly over the spectrum with an average of
three features per \AA .

Since fixed pattern features are typically shallow and tend to be slightly broader than atomic features, they are not as easily confused 
with weak atomic absorption.  However, they can easily mimic a sharper diffuse band.  
The question then stands whether it is possible to 
define criteria that can distinguish fixed pattern noise from interstellar features.  Distinguishing between 
real and false features is of course nearly impossible using a single exposure.  To be confident of the existence
of a feature, it is best to confirm the absorption in at least two observations with different central wavelengths.
Unfortunately, the only significant amount of overlap in the high S/N X Per data is between 1320 \AA\ 
and 1440 \AA , significantly limiting the region of usable data.  We can also use the regions 
of overlap between echelle orders in a single observation.  We set out to distinguish real interstellar from
fixed pattern noise features by applying the following criteria: 1) features must be measured to at least 2$\sigma$ 
significance in both observations, 2) features must be consistent within 2$\sigma$ across observations in their
measured width, depth, and central wavelength, and 3) the measured significance of the coadded feature must be greater 
than the measured significance of the feature in either individual observation.  The third criterion is somewhat 
redundant to the other two; however, in practice it was still useful in eliminating some cases.  For this third 
criterion we compute a significance ratio:
\begin{equation}
\xi =  \frac{\sigma _{single} W_{\lambda coadd}}{\sigma _{coadd} W_{\lambda {single}}}
\label{eqn_sigratio}
\end{equation}
where $W_{\lambda}$ and $\sigma$ are the measured equivalent width and error respectively.

To test the effectiveness of these criteria, we ran several tests on various synthetic data sets.  First 
we ran 5000 simulations by coadding two randomly generated datasets where
either both datasets contained a 3$\sigma$ significant Gaussian feature (representative of a true interstellar feature) 
or only one dataset contained a 3$\sigma$ significant Gaussian feature (representative of an instrumental feature).  
In simulations where the S/N of individual data sets were equal, we found that our criteria 
with a cutoff of $\xi$ $>$ 1.0 rejected more than 99.9\% of instrumental features (type I errors) while retaining 
the majority of real features (type II errors).  Thus, for truly isolated features (features with no possibility 
of coincidental overlap with other features) our criteria are sufficient to reject instrumental features at high accuracy.
However, the bigger problem is that the high occurrence of instrumental features in STIS echelle datasets 
makes the chance of coincidental overlap of instrumental features in two different observations 
by no means negligible.

We can estimate roughly the number of instrumental features that would statistically coincide with one another through
some simple calculations.  As mentioned previously, we estimate that on average there are 3 statistically significant 
features per \AA\ in the X Per data.  Standard errors on wavelength measurements are usually of order 0.01 \AA\ 
(taking into account the $\sim$ 1 pixel accuracy of STIS wavelength calibrations).  Thus, the probability that 
an instrumental feature in one spectrum will coincide with an instrumental feature in another spectrum 
may be about 12\%.
On average for features that are a few sigma significant, the error in widths and depths are 
usually 10\% to 30 \% of their best fit values.  If we use the distribution of instrumental feature FWHMs
and depths found in scanning the X Per dataset as rough probability mass functions, we can run a simple discreet 
Monte Carlo simulation to estimate the probability that two features will coincidentally be consistent within 
2$\sigma$.  We find that the probability of two features FWHMs coinciding to be roughly 40\%
while the probability of two features depths coinciding to be roughly 60\%.

If we treat the instrumental features width, depth, and wavelength as independent 
(not a bad assumption given the lack of correlation) we thus might expect of order 10
($0.12\times0.40\times0.60\times360$)
false detections passing our criteria due to coincidental overlap of fixed pattern noise.
This is of course very concerning.
With any number of expected false detections the $P(Real|Pass)$ will be too low to 
confidently claim a real interstellar feature.  We confirmed this concern by generating several datasets 
that had features with $\le$ 3\% depths randomly distributed through them.  We then scanned these spectra for 
features and found how many passed our criteria when two synthetic datasets were compared.  In all cases a few
features would pass all criteria.  This confirmed that, given the described criteria, we could not sufficiently 
reject instrumental features having only two overlapping datasets.

Thus, for the current study using archival data we find that it is not possible to confidently claim a detection
of an unidentified feature if that feature has characteristics similar to the fixed pattern noise.  Improvements 
can be made if observations are planned more carefully to result in more overlapping observations; however, for the 
current dataset this problem of fixed pattern noise implies that we must limit our search to features that do not have
similar characteristics to the fixed pattern noise.  From Figure \ref{fig_fpchar} it is clear that 
the easiest way to exclude the majority of fixed pattern noise features is to have a minimum depth cutoff for real features.
Requiring features to have depths $>$ 3\% excludes 95\% of instrumental features detected in the STIS data.  With such a 
cutoff, the expected number of false detections in 120 \AA\ of X Per spectra is much less than one.

\subsection{Verifying Features in Other Sightlines}
For all unidentified interstellar features detected in X Per, several other sightlines were 
checked to attempt to verify the existence of the unknown feature.  Sightlines were chosen 
from those included in previous interstellar medium studies (Burgh, France, \& McCandliss 2007; 
Sonnentrucker et al. 2007, Destree, Snow, \& Black 2009) and those that had STIS data in the 
region of the feature.

In verifying features in other sightlines, we are only looking for a feature at a known wavelength; 
however, it is still important to assess whether instrumental features will pose a problem in verifying 
interstellar features.  If we were to have no criteria for detecting a feature in other sightlines other than 
its wavelength and width being consistent within errors, the probability that a weak instrumental feature 
would have the correct wavelength and width in any given sightline would be about 5\% ($0.12\times0.4$).  
However, given that we checked about 20 sightlines in attempting to verify interstellar features, 
there is a non-negligible probability that a weak instrumental feature will have the correct wavelength
and width in at least one of our sightlines.  Thus, we additionally require that any verification also 
have a depth of at least 3\% of the continuum level.

\section{Results and Discussion}
The blind search resulted in 527 initial features being flagged between 1320 \AA\ and 1440 \AA\ 
in the coadded spectrum, before any cuts were applied.  The overlap region between echelle orders was also 
used to look for unknown features and 27 initial features were flagged in these regions.
Of these 554 features, we assigned 100 features to known absorption lines.  In a detailed look through the data
 and the list of flagged features between 1320 \AA\ and 1440 \AA , we did not find any significant, isolated, known features
that were not flagged by the blind search.  Of those that were not identified as known features,
only two were confirmed as likely true interstellar absorption features through the criteria explained previously.  
The vast majority of features were rejected due to not meeting cutoff criteria (too wide or too shallow profile), or 
poorly defined continuum (as discussed previously).

\subsection{1297 \AA\ and 1300 \AA\ Features}
The two unidentified features (1297.21 \AA\ and 1300.45 \AA ) were 6$\sigma$ and 9$\sigma$ significant 
respectively, when fit with a Gaussian profile. Measured equivalent widths for both features are 
listed in Table \ref{tab_features}.  The rest wavelengths of both features were based on the best 
fit central wavelength and the velocity of nearby neutral sulfur lines.
Though we do not know the velocity structure of the unidentified lines,
it is reasonable to assume that the unknown carriers generally follow the 
spacial distribution of other neutral and ionic species.  S I lines tend to 
be easy to measure and trace colder diffuse interstellar clouds well.

\begin{figure}
\begin{center}
\scalebox{.8}[.8]{\includegraphics[angle=0]{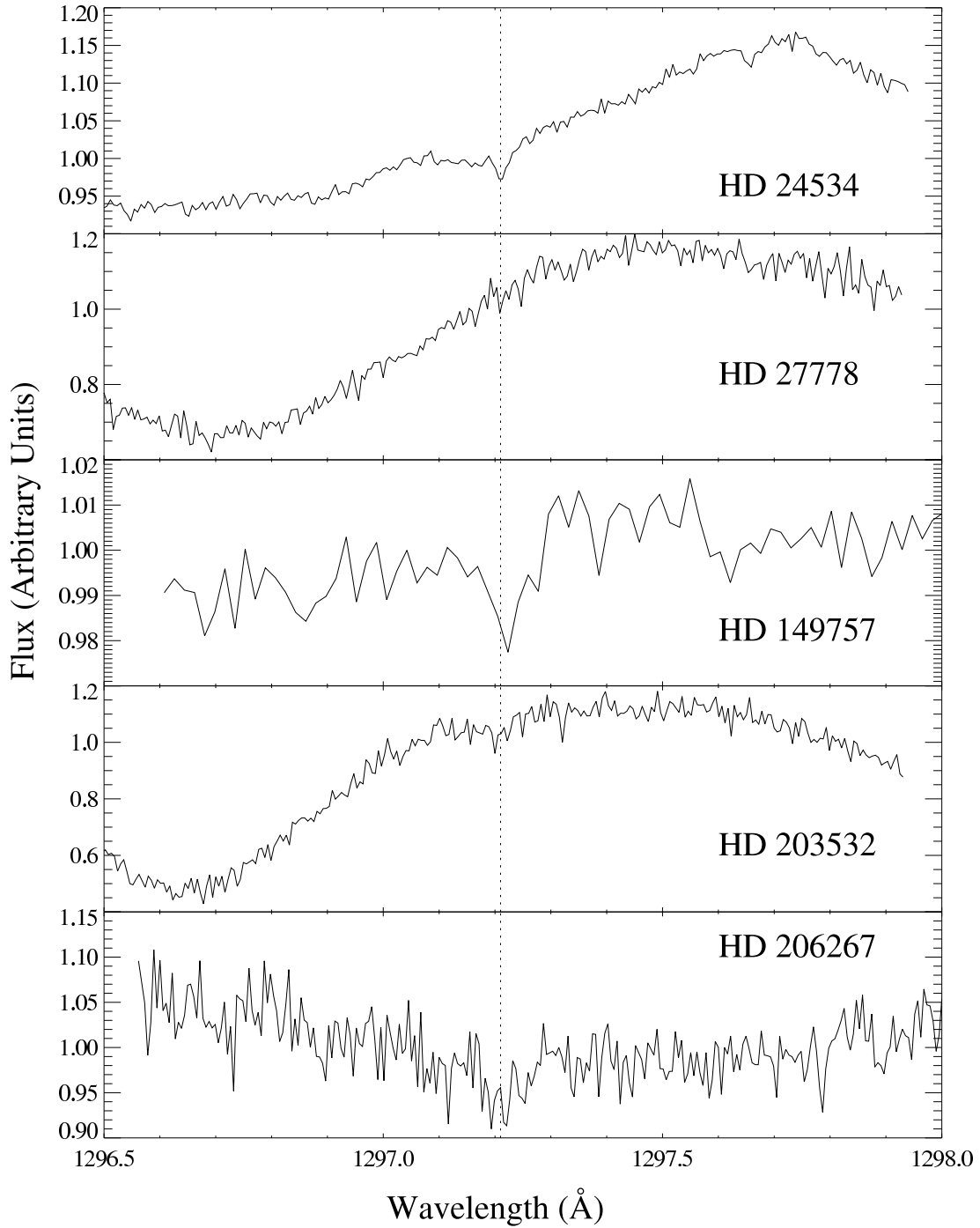}}
\caption{Plot of 1297 \AA\ feature in 5 sightlines.  The 1297 \AA\ feature was significantly detected in the sightlines to HD 24534, HD 203532 and HD 206267.  Dotted line marks the mean wavelength (rest wavelength based on nearby S I lines). \label{fig_1297}}
\end{center}
\end{figure}

\begin{figure}
\begin{center}
\scalebox{.8}[.8]{\includegraphics[angle=0]{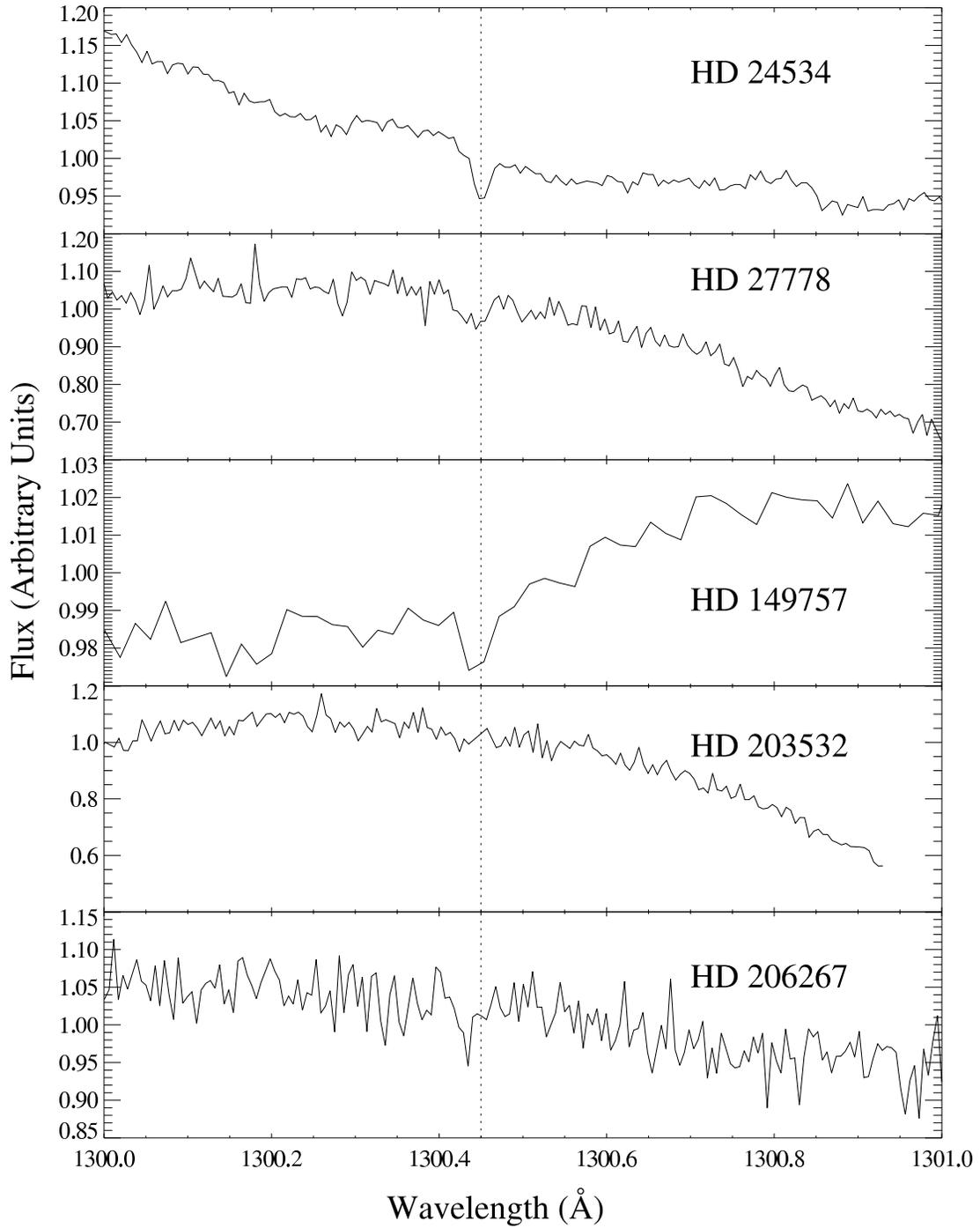}}
\caption{Plot of 1300 \AA\ feature in 5 sightlines.  The 1300 \AA\ feature was significantly detected in the sightlines to HD 24534 and HD 27778.  Dotted line marks mean wavelength (rest wavelength based on nearby S I lines). \label{fig_1300}}
\end{center}
\end{figure}

The feature at 1297.21 \AA\ has an uncertain profile --
in X Per it seems to have primarily one narrow feature (FWHM of 5 to 6 km/s); however, in trying to 
verify the feature in other sightlines, it appears to 
perhaps have wider band structure as well (see Figure \ref{fig_1297}).
In contrast, the feature at 1300.45 \AA\ is clearly narrow, having a FWHM of about 5 km/s in X Per (see Figure \ref{fig_1300}).  
This width is comparable to other atomic features like O I or C I and is wider than many of the sharp CO lines.

An important aspect of studying these features is to attempt to verify their 
existence in other sightlines as well.  In order to attempt this, we must make some assumptions 
about the velocity structure of the unidentified features.
We again used nearby S I lines to shift spectra to their rest wavelengths.
From all STIS datasets searched, including X Per,
the 1297.21 \AA\ feature was detected in three sightlines (see Table \ref{tab_1297}) 
while the 1300.45 \AA\ feature was detected in two (see Table \ref{tab_1300}).
Some of the high S/N GHRS data sets also showed possible absorption for both 
of the features identified in X Per. These features are shown in Figures \ref{fig_1297} 
and \ref{fig_1300}; however, given their very weak strength 
and our not having investigated GHRS data more thoroughly, we do not claim
these as confident detections in the tables.

\subsection{Feature Identifications}
Both features were checked against published compilations of 
atomic, C2, CH, CH$_2$, CH$_3$, CO, CN, CS, H$_2$, H$_2$O, HCl, OH, NO$^+$, and 
SiO lines (see Table \ref{tab_lines}).
The wavelengths of the detected features were also compared to online databases
of atomic and molecular transitions at
the National Institute of Standards and Technology\footnote{\tt http://physics.nist.gov/PhysRefData/ASD/},
the Harvard-Smithsonian Center for Astrophysics\footnote{\tt http://www.cfa.harvard.edu/amp/ampdata/cfamols.html},
and the Max Planck Institute for Chemistry in Mainz\footnote{\tt http://www.atmosphere.mpg.de/enid/2295}
as well as large published compilations of molecular transitions (Herzberg 1966, and Huber \& Herzberg 1979).
In all these lists, no atomic or molecular lines were found to be
compelling identifications of either the 1297.21 \AA\ or 1300.45 \AA\ features.

A few wavelengths of molecular lines did coincide with the detected features; however,
all of these cases were rejected.  The 1297.21 \AA\ feature had three different molecular transitions 
in close proximity to it. First, a line of ethylene oxide (C$_2$H$_4$O)
at 1297.2 \AA\ matched the wavelength of the feature in X Per nearly exactly.  This 
molecule was ruled out as a carrier, though, as other stronger lines of ethylene oxide
were not observed.  Second, a line of DCl also fell at 1297.2 \AA .  This is an extraordinarily
unlikely carrier as the C-X band HCl was not significantly detected in the X Per data and
the abundance of DCl should be orders of magnitude smaller than HCl.  Third, a weak line of
CH$_3$ has been observed in the lab at 1297.2 \AA ; however, we reject this as a
possible identification because this line is relatively weak while stronger lines of CH$_3$ are not 
observed in the spectrum of X Per.
Regarding the 1300.45 \AA\ feature, the I-X(13,0) band of CO has a rest wavelength of
1300.58 \AA , but this is a very unlikely identification as the I-X system is completely forbidden 
in a non-rotating molecule and can only occur weakly in high j levels due to Coriolis interaction 
(Herzberg et al. 1966).

\subsection{Upper Limits}
Glancing at these two best cases reveals a few interesting facts.  First, we detected no  
strong features; both absorption lines equivalent widths are under 2 m\AA .  In one way 
this result might be expected, as stronger unidentified features probably would have been reported 
before now.  On the other hand, it is somewhat surprising that there is so little absorption by 
molecules we cannot identify.  The visible DIBs indicate to us that a significant amount of 
unknown carriers exist in most interstellar clouds, yet corollaries to the carriers that produce the visible DIBs with 
equivalent widths of up to several hundred m\AA\ have nearly no narrow absorption in the 
UV region we studied.  Because of the limited extent of this study there are several possible explanations 
for the lack of features.  First, larger molecules like the DIB carriers may have few narrow transitions in 
the region we studied.  Second, in X Per, the dominant ionization state of organics may not favor narrower 
features in the UV.  Third, because of the large number of possible PAHs and other organics, the column 
densities of any single species may not be great enough to create strong UV features.
We suspect that the wavelength region used (as dictated by existing, overlapping 
archival data) may be one of the main reasons no DIB-like features were found in the current study.
The near-UV between 2500 \AA\ and 3000 \AA\ may be a better place to search for small neutral PAHs (F. Salama, private
communication).

\begin{figure}
\begin{center}
\scalebox{.75}[.75]{\includegraphics[angle=0]{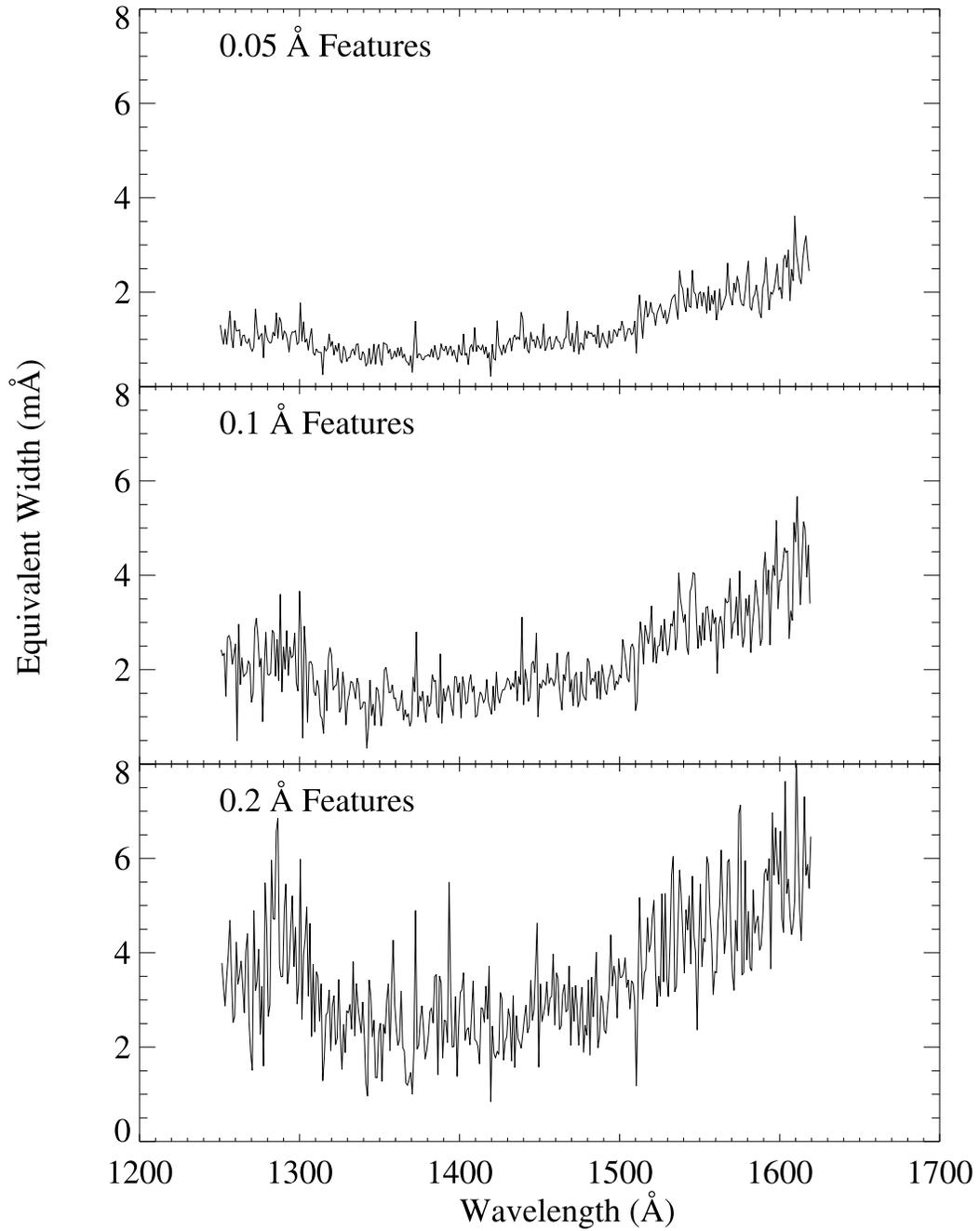}}
\caption{Plots of 2$\sigma$ upper limits on unidentified features in X Per data when integrating over segments with widths of 0.05 \AA\ (top), 0.1 \AA\ (middle), and 0.2 \AA\ (bottom). \label{fig_uls}}
\end{center}
\end{figure}

Even with the influence of STIS instrumental features, we can still set upper limits on the size of the 
features not detected. To do this we scanned the entire STIS far UV spectrum of X Per integrating over 
0.05 \AA , 0.10 \AA , and 0.20 \AA\ regions (approximately 8, 16 and 33 pixels respectively) 
while ignoring known features.  In Figure \ref{fig_uls} we show the 2$\sigma$ upper 
limits across the spectrum.  The shapes of the 
upper limit curves are due almost exclusively to the varying S/N across the spectrum.
One will notice quickly that for an area of over 100 \AA , we can set very stringent upper limits of 
less than 1 m\AA\ for narrow features.  This again confirms our previous argument that there are simply not 
strong unidentified narrow features in the far UV spectrum of X Per.

\subsection{Previous Unidentified Features}
One last exercise performed with the X Per dataset was to look for previously reported 
unidentified features (UIDs).  Over the history of ISM research using UV spectroscopy several
studies have reported unidentified lines in the UV spectrum (see  Table \ref{tab_previousuids}).  
We attempted to verify or set upper limits on 
any previously reported UIDs in the STIS range in the spectrum of X Per.  UIDs were taken from four studies and are 
listed in Table \ref{tab_previousuids} along with X Per equivalent width measurements.  Previously reported 
UIDs that have been identified were not included in the table.

Of the 18 features in the STIS far UV range, four had features in X Per that were relatively close 
in wavelength (1248 \AA , 1253 \AA , 1384 \AA , and 1490\AA ).  However, all four of the features
in X Per were relatively weak and had depths $<$ 3\% of the continuum level.  Also three of the 
four features could not be verified in more than one observation.  Thus, it is likely that three of the
features (1248 \AA , 1253 \AA , and 1490\AA ) are simply due to fixed pattern noise.  The feature at 
1384 \AA\ can be verified in two different observations of X Per; however, because of the feature's shallow 
depth (2.3\% of the continuum) and broad width (FWHM $\sim$ 120 m\AA ) it could easily be 
a stellar or fixed pattern noise feature.  Thus for all the previously reported UIDs whose 
wavelengths fall within the range of the X Per dataset, we find no convincing 
corresponding interstellar feature in the sightline to X Per.

\section{Conclusions and Future Work}
In closely analyzing a small section  of high resolution STIS data, we report two 
features that remain unidentified.  The profiles of both unidentified lines are relatively narrow
with widths similar to atomic lines and therefore should probably not be thought of as UV DIBs.
The main factors limiting the current study were interference from 
shallow ($\leq$ 3\% depth) instrumental features and insufficient overlapping observations to accurately distinguish 
interstellar from fixed pattern noise features.  In seeking to verify previously reported UIDs, 
we find no convincing interstellar features in the X Per data.

The results from our study of STIS echelle characteristics should impart significant caution to any 
future measurements of features in STIS echelle data that are very shallow in nature.
More generally, any studies that seek to detect weaker features at unknown wavelengths or velocities
could benefit from a more thorough investigation of instrumental effects and a more rigorous 
estimation of the confidence level of any given detection.  As is clear from this study, a single
feature in one observation is by no means sufficient, and, if instrumental effects are bad enough, even 
two independent observations of the same feature may not be sufficient to confidently claim a detection.

Overall, it is somewhat surprising that there is so little absorption in the far UV from unidentifiable carriers.
Many molecules have ground state transitions within the STIS far UV bandpass.  It may be that abundances
for most molecules are too low in the diffuse ISM to readily detect them.  Yet, this is surprising 
in part when considering the strength of some of the visible DIBs.

There are many ways to extend the current work.  The archive of medium and high resolution
data is vastly untouched in many respects.  Few comprehensive surveys have been done with 
newer datasets.  Most previously reported UIDs resulted from studies of $\zeta$ Oph.
One promising prospect would be to attempt to stack STIS observations of multiple interstellar sightlines,
thereby increasing the S/N of a coadded spectrum and also minimizing the effects of instrumental
blemishes.  The search for unidentified features should be extended into the near UV,
as organics like small neutral PAHs are expected to have transitions in this region.

Another clear possibility for future work is searching for still larger features through modeling the 
stellar spectrum of X Per (or other sightlines) and dividing such a model from the STIS data.  
Such modeling was not attempted 
in the present search as the current analysis focuses on narrow features overlooked in previous studies.
Wider features will be very difficult to investigate as even the best models tend to still have 
significant discrepancies from observed spectra due to the inherent complexity of hot stars' UV spectra.
Even with such discrepancies and difficulties, such a study is warranted as stellar models and computing 
ability continue to improve.

Lastly, with the coming launch of the Cosmic Origins Spectrograph (COS), probing much denser 
regions of the ISM will be possible and a greater number of molecular lines may be readily observable.
Future COS and STIS observations need to be planned to result in three (or preferably more) 
high S/N exposures that all overlap over a large section of the UV (using slightly different central wavelength 
settings or FP-positions).  If such data were available, much more could be done to discriminate interstellar from 
instrumental features and to push further into the unknown of the interstellar UV spectrum.

\acknowledgments
This research was supported by NASA grant NNX08AC14G.
We are grateful to Kevin France, John Black, Farid Salama, Eric Burgh, Adam Jensen, Steve Penton, Steve Federman, 
Charles Proffitt, Tom Ayres, and Charles Danforth for their insightful comments and input.  
We want to give a special thanks to Rachel Destree for her continual assistance in editing our publications.
All of the data presented in this paper were obtained from the Multimission Archive 
at the Space Telescope Science Institute (MAST). STScI is operated by the Association 
of Universities for Research in Astronomy, Inc., under NASA contract NAS5-26555. 
Support for MAST for non-HST data is provided by the NASA Office of Space Science 
via grant NAG5-7584 and by other grants and contracts.



\clearpage

\begin{deluxetable}{lr}
\tablecolumns{2}
\tablewidth{0pc}
\tabletypesize{\footnotesize}
\tablecaption{Published line lists used in identifying interstellar features \label{tab_lines}}
\tablehead{Species & Reference}
\startdata
Atomic	& Morton (2000)	\\
''	& Morton (2003)	\\
C$_2$	& Sonnentrucker et al. (2007)	\\
CH	& Herzberg \& Johns (1969)	\\
''	& Watson (2001)	\\
CH$_2$	& Herzberg (1961)	\\
CH$_3$	& Herzberg (1961)	\\
CO	& du Plessis, Rohwer, \& Steenkamp (2006)	\\
''	& Eidelsberg \& Rostas (2003)	\\
''	& Morton \& Noreau (1994)	\\
''	& Herzberg et al. (1966)	\\
CN	& Lutz (1970)	\\
CS 	& Destree, Snow, \& Black (2009)	\\
''	& Stark, Yoshino, \& Smith (1987)	\\
H$_2$	& Abgrall et. al (2003a)	\\
''	& Abgrall et. al (2003b)	\\
H$_2$O	& Price (1936)	\\
''	& Watanabe \& Zelikoff (1953)	\\
HCl	& Tilford, Ginter, \& Vanderslice (1970)	\\
OH	& Douglas (1974)	\\
''	& Nee \& Lee (1984)	\\
NO$^+$	& Alberti \& Douglas (1975)	\\
SiO	& Lagerqvist, Renhorn, \& Elander (1973)	\\
	
\enddata
\end{deluxetable}

\begin{deluxetable}{lcccc}
\tablecolumns{5}
\tablewidth{0pc}
\tabletypesize{\footnotesize}
\tablehead{$\lambda$	& FWHM  &	W$_{\lambda\ 1}$	& W$_{\lambda\ 2}$	& W$_{\lambda\ coadd}$ \\
(\AA )	&(m\AA )	&(m\AA )	&(m\AA )	&(m\AA )}
\tablecaption{Unidentified absorption features in the spectrum of X Per \label{tab_features}}
\startdata
1297.21	& 24	& 1.1 $\pm$ 0.2		& 0.75 $\pm$ 0.18	& 0.86 $\pm$ 0.15		\\ 
1300.45	& 20	& 1.4 $\pm$ 0.2		& 1.2 $\pm$ 0.2		& 1.2 $\pm$ 0.1		\\ 
\enddata
\end{deluxetable}

\begin{deluxetable}{lrrrr}
\tablecolumns{5}
\tablewidth{0pc}
\tabletypesize{\footnotesize}
\tablecaption{Detections of the 1297 \AA\ feature \label{tab_1297}}
\tablehead{Sightline & W$_{1297}$ & $\lambda_{observed}$ & S I velocity & $\lambda_{rest}$\tablenotemark{a} \\
 & (m\AA )	& (\AA )	& (km/s)	&(\AA )}
\startdata
HD 24534	&0.86 $\pm$ 0.15&1297.27	&14.8	&1297.21	\\
HD 203532	&4.9 $\pm$ 1.5	&1297.26	&14.8	&1297.20	\\
HD 206267	&3.9 $\pm$ 1.1	&1297.15	&-16.6	&1297.22	\\
Mean		&		&		&	&1297.21 $\pm$ 0.01 \\
\enddata
\tablenotetext{a}{Rest wavelengths were calculated using the velocity of nearby S I lines}
\end{deluxetable}

\begin{deluxetable}{lrrrr}
\tablecolumns{5}
\tablewidth{0pc}
\tabletypesize{\footnotesize}
\tablecaption{Detections of the 1300 \AA\ feature \label{tab_1300}}
\tablehead{Sightline & W$_{1300}$ & $\lambda_{observed}$ & S I velocity & $\lambda_{rest}$\tablenotemark{a} \\
 & (m\AA )	& (\AA )	& (km/s)	&(\AA )}
\startdata
HD 24534	&1.4 $\pm$ 0.2	&1300.51	&14.8	&1300.45	\\
HD 27778	&2.4 $\pm$ 0.9	&1300.51	&15.5	&1300.44	\\
Mean		&		&		&	&1300.45 $\pm$ 0.01 \\
\enddata
\tablenotetext{a}{Rest wavelengths were calculated using the velocity of nearby S I lines}
\end{deluxetable}

\begin{deluxetable}{lccccc}
\tablecolumns{6}
\tablewidth{0pc}
\tabletypesize{\footnotesize}
\tablehead{$\lambda$  & Instrument&W$_{\lambda}$ &Reference\tablenotemark{b}  & X Per & X Per \\
(\AA ) &  &(m\AA ) &  & $\lambda$ (\AA )	& W$_{\lambda}$ (m\AA )}
\tablecaption{Previously reported UIDs in the STIS far-UV range\tablenotemark{a} \label{tab_previousuids}}
\startdata
1156.27	& {\it Copernicus}	& 12.1	& 1	& ...		& ...	\\
1159.37	& {\it Copernicus}	& 5.4	& 1	& ...		& ...	\\
1178.20	& {\it IUE}	& 36.9	& 2		& ...		& ...	\\
1192.20	& {\it IUE}	& 37.1	& 2		& ...		& ...	\\
1229.84	& GHRS	& 3.9	& 3		& ...		& ...	\\
1243.15	& {\it IUE}	& 22.2	& 2		& ...		& ...	\\
1248.07	& {\it IUE}	& 16.9	& 2		& 1248.05	& 0.7$\pm$0.3\tablenotemark{c}	\\
1249.9	& {\it IUE}	& 17.6	& 2		& ...		&$\leq$4.6	\\
1252.77	& GHRS	& 0.9	& 3		& 1252.66	& 2.9$\pm$0.5\tablenotemark{c}	\\
1308.09	& {\it Copernicus}	& 9.2	& 1	& ...		&$\leq$0.6	\\
1346.54	& GHRS	& 1.73	& 4		& ...		&$\leq$0.3	\\
1383.45	& {\it IUE}	& 6.6 	& 2		& ...		&$\leq$2.5	\\
1383.92	& {\it IUE}	& 6.8 	& 2		& ...		&$\leq$1.1	\\
1384.27	& {\it IUE}	& 11.4	& 2		&1384.11	&3.1$\pm$0.4\tablenotemark{d}	\\
1404.8	& GHRS	& 0.2	& 3		& ...		&$\leq$2.2	\\
1424.24	& GHRS	& 0.45	& 3		& ...		&$\leq$1.3	\\
1490.1	& {\it IUE}	& 8.9 	& 2		&1490.12	&1.5$\pm$0.5\tablenotemark{c}	\\
1538.0	& {\it IUE}	& 21.2	& 2		&...		&$\leq$5.2\tablenotemark{e}	\\

\enddata

\tablenotetext{a}{All studies are of the sightline $\zeta$ Oph }
\tablenotetext{b}{References: 1) Morton 1978, 2) Pwa \& Pottasch 1986, 3)Brandt et al. 1996, and 4)Federman et al. 1995}
\tablenotetext{c}{Feature is highly uncertain as it is shallow ($<$3\% depth) and cannot be verified in multiple observations}
\tablenotetext{d}{Feature is uncertain as it is shallow ($<$3\% depth)}	
\tablenotetext{e}{Uncertain continuum}	

\end{deluxetable}


\begin{thebibliography}{}


\bibitem[]{} Abgrall, H., Roueff, E., Launay, F., Roncin, J. Y., \& Subtil, J. L. 1993, A\&AS, 101, 273
\bibitem[]{} Abgrall, H., Roueff, E., Launay, F., Roncin, J. Y., \& Subtil, J. L. 1993, A\&AS, 101, 323
\bibitem[]{} Alberti, F. \& Douglas, A. E. 1975, CaJPh, 53, 1179
\bibitem[]{} Andr\`{e}, M. K. et al. 2003, ApJ, 591, 1000
\bibitem[]{} Brandt, J. C. et al. 1996, AJ, 112, 1128
\bibitem[]{} Burgh, E. B., France, K., \& McCandliss, S. R. 2007, ApJ, 658, 446
\bibitem[]{} Clayton, G. C. et al. 2003, ApJ, 592, 947
\bibitem[]{} Destree, J. D., Snow, T. P. \& Black, J. H. 2009, ApJ in press
\bibitem[]{} Draine, B. T. 2003, ARA\&A, 41, 241
\bibitem[]{} Douglas, A. E. 1974, CaJPh, 52, 318
\bibitem[]{} du Plessis, A., Rohwer, E. G., \& Steenkamp, C. M. 2006, ApJS, 165, 432
\bibitem[]{} Eidelsberg, M. \& Rostas, F. 2003, ApJS, 145, 89
\bibitem[]{} Federman, S. R., Cardell, J. A., van Dishoeck, E. F., Lambert, D. L., Black, J. H. 1995, ApJ, 445, 325
\bibitem[]{} Herzberg, G. 1961, RSPSA, 262, 291 
\bibitem[]{} Herzberg, G. 1966, Molecular Spectra and Molecular Structure, Vol. 3 (Princeton, NJ: D. Van Nortrand Company)
\bibitem[]{} Herzberg, G., Simmons, J. D., Bass, A. M., \& Tilford S. G. 1966, CaJPh, 44, 3039
\bibitem[]{} Herzberg, G. \& Johns, J. W. C. 1969, ApJ, 158, 399
\bibitem[]{} Huber, K. P \& Herzberg, G. 1979, Molecular Spectra and Molecular Structure, Vol. 4 (New York, NY: D. Van Nortrand Reinhold Company)
\bibitem[]{} Lagerqvist, A., Renhorn, I., \& Elander, N. 1973, JMoSp, 46, 285 
\bibitem[]{} Lanz, T., \& Hubeny, I. 2003, ApJS, 146, 417
\bibitem[]{} Lanz, T., \& Hubeny, I. 2007, ApJS, 169, 83
\bibitem[]{} Leger, A. \& Puget, J. L. 1984, A\&A, 137, 5
\bibitem[]{} Lutz, B. L. 1970, CaJPh, 48, 1192
\bibitem[]{} Morton, D. C. 1978, MNRAS, 184, 713
\bibitem[]{} Morton, D. C. 2000, ApJS, 130, 403
\bibitem[]{} Morton, D. C. 2003, ApJS, 149, 205
\bibitem[]{} Morton, D. C. \& Noreau, L. 1994, ApJS, 95, 301
\bibitem[]{} Nee, J. B. \& Lee, L. C. 1984, JChPh, 81, 31
\bibitem[]{} Penny, L. R. 1996, ApJ, 463, 737
\bibitem[]{} Price, W. C . 1936, JChPh, 4, 147  
Pwa, T. H. \& Pottasch, S. R. 1986, A\&A, 164, 116
\bibitem[]{} Salama, F. 2007, in Molecules in Space and Labaratory, ed: J.L. Lemaire \& F. Combes. (Paris, S. Diana), 51
\bibitem[]{} Salama, F., Bakes, E. L. O., Allamandola, L. J., Tielens, A. G. G. M. 1996, ApJ, 458, 621
\bibitem[]{} Seab, C. G., \& Snow, T. P. 1985, ApJ, 295, 485
\bibitem[]{} Sonnentrucker, P., Welty, D. E., Thorburn, J. A., \& York, D. G. 2007, ApJS, 168, 58
\bibitem[]{} Snow, T. P. \& Bierbaum, V. M. 2008, Annual Reviews of Analytical Chemistry, 1, in press
\bibitem[]{} Snow, T. P., York, D. G., \& Resnick, M. 1977, PASP, 89, 758
\bibitem[]{} Stark, G., Yoshino, K., \& Smith, P. L. 1987, JMoSp, 124, 420
\bibitem[]{} Tilford, S. G., Ginter, M. L., \& Vanderslice, J. T. 1970, JMoSp, 33, 505
\bibitem[]{} Thorburn, J. A. et al. 2003, ApJ, 584, 339
\bibitem[]{} Tripp, T. M.,  Cardelli, J. A., \& Savage, B. D. 1994, AJ, 107, 645
\bibitem[]{} Watanabe, K. \& Zelikoff, M. 1953, JOSA, 43, 753
\bibitem[]{} Watson, J. K. G. 2001, ApJ, 555, 472



\end{thebibliography}
\end{document}